\documentstyle[12pt,epsfig]{article}

\newcounter{subequation}[equation]
\makeatletter
\expandafter\let\expandafter\reset@font\csname reset@font\endcsname
\newenvironment{subeqnarray}
  {\arraycolsep1pt
    \def\@eqnnum\stepcounter##1{\stepcounter{subequation}{\reset@font\rm
      (\theequation\alph{subequation})}}\eqnarray}%
  {\endeqnarray\stepcounter{equation}}
\makeatother

\newcommand{\labelcaption}[2]{\caption[#1]{\label{#1}#2}}
\newcommand{\bu}{{\bar{U}}}
\newcommand{\bb}{{\bar{B}}}

\begin{document}

\hbox to\hsize{%
  \hfil\vbox{%
       \hbox{gr-qc/9708036}%
       \hbox{BUTP-97/23}%
       \hbox{August 17, 1997}%
       }}
\vspace{5mm}
\begin{center}
\LARGE\bf
Non-Abelian black holes: 

The inside story
\footnote{Talk given at the international Workshop on 
The Internal Structure of Black Holes and Spacetime Singularities, 
Haifa, Israel, June 29 -- July 3, 1997}

\vskip5mm
\large Peter Breitenlohner$~{^\S}$,
George Lavrelashvili$~{^\ddagger}$
\footnote{On leave of absence from Tbilisi
Mathematical Institute, 380093 Tbilisi, Georgia}\\
{\normalsize and} Dieter Maison$~{^\S}$

\vspace{3mm}
{\small\sl
Max-Planck-Institut f\"ur Physik$~{^\S}$\\
--- Werner Heisenberg Institut ---\\
F\"ohringer Ring 6\\
80805 Munich (Fed.\ Rep.\ Germany)

\vspace{3mm}
Institute for Theoretical Physics$~{^\ddagger}$\\
University of Bern\\
Sidlerstrasse 5 \\
CH-3012 Bern, Switzerland\\}

\end{center}

\begin{center}\bf Abstract\end{center}
\vspace{1mm}\noindent
Recent progress in understanding of the internal structure of 
non-Abelian black holes is discussed.

\newpage

\section{Introduction}

Let us start recalling  
what is known about the non-Abelian black holes in general.
The non-Abelian story begun in 1988 when Bartnik and McKinnon (BK)
unexpectedly found  \cite{lav-BK} numerically a discrete sequence 
of globally regular, particle like solutions of the 
Einstein--Yang--Mills (EYM) theory.  
Soon the same model was solved with the different boundary conditions 
corresponding to black holes \cite{lav-bholes}.
Numerical findings were confirmed by mathematically rigorous 
existence proofs \cite{lav-existence1,lav-BFM} of both
regular and black hole solutions.
It turned out that all the above solutions of the EYM theory
are classically unstable against small perturbations.
In addition to the genuine gravitational instabilities \cite{lav-SZ} 
there also instabilities of topological origin \cite{lav-LM3,lav-VBLS} 
related to the sphaleron nature of the solitons \cite{lav-VG,lav-GLzero}.

A few related systems were investigated. 
It was shown \cite{lav-LM1,lav-BIZ1}
that the gravitational field can be replaced by a dilaton, so that the   
YM-dilaton theory in flat space has a tower of solutions 
similar to the BK sequence. 
The combined EYMD theory \cite{lav-LM2,lav-BIZ2,lav-DG1,lav-TM,lav-KKS}
was  shown to possess both regular and black hole solutions
for any value of the dilaton coupling constant. 
For the EYMH theory with a Higgs doublet \cite{lav-GMN}
it was shown that the theory 
in addition to the gravitating sphaleron solution contains  
its BK type excitations.  

Furthermore the EYMH theory with a triplet Higgs \cite{lav-BFMmonopole,lav-LNW}
was studied. This theory is interesting since in the flat 
limit it contains t`Hooft-Polyakov monopoles, 
which are known to be stable. 
It was shown that the basic monopoles continue to exist,
when gravity is switched on, at least as long as the gravitational
self-interaction is not too strong. In addition 
the monopole admits unstable BK type excitations. 

Adding a cosmological constant to the EYM theory one obtains
non-asymptotically flat analogues of the BK solutions 
\cite{lav-TMT,lav-VSLHB}. 

Whereas in the above study the globally regular solutions were 
completely analyzed, the investigation of the black hole solutions 
was not complete since their internal structure was unknown.
Recently this problem was investigated 
independently by us \cite{lav-BLM} and by 
Donets, Gal'tsov and Zotov \cite{lav-DGZ1}. 
Our main results on the EYM case essentially agree with theirs, 
although we differ in some details. 

We found it adequate to describe  the generic behavior as a kind 
of mass inflation closely related to the  ``usual'' mass inflation 
(see e.g. \cite{lav-PI,lav-Ori1,lav-Page,lav-Ori2,lav-BI}
and many references in the present proceedings).

In addition to the generic behaviour there are three different 
types of special solutions,
which are obtained by fine tuning of the initial data at the 
horizon. There are solutions with Reissner-Nordstr{\o}m (RN), 
Schwarschild and pseudo-RN type behaviour.   

The present contribution is essentially based on our paper \cite{lav-BLM}. 

\section{Field Equations}\label{chapans}

The action of the EYMH theory  is 
\begin{equation}\label{Action0}
S=\frac{1}{4\pi}\int\Bigl(-\frac{1}{4 G}R-{1\over 4g^2}F^2+
{1\over 2} |D_\mu\Phi|^2-V(\Phi)\Bigr)\sqrt{-g}\;d^4x\;,
\end{equation}
where \ $g$ denote the gauge coupling constant, 
$G$ is Newton's constant, $F$ is the field strenght of the  
{\sl SU(2)\/} Yang-Mills field
and $V(\Phi)$ is the usual quartic Higgs potential.
The pure EYM action and corresponding equations can be
trivially obtained from the EYMH ones by putting 
the Higgs field $\Phi$ and its potential $V(\Phi)$ to zero. 

For the static, spherically symmetric metric we use the parametrization
\begin{equation}\label{Metr}
ds^2=A^2 B dt^2-{dR^2\over B}-r^2(R)d\Omega^2\;,
\end{equation}
with $d\Omega^2=d\theta^2+\sin^2\theta d\varphi^2$
and three independent functions $A$, $B$, $r$ of a radial coordinate
$R$, which has, in contrast to $r$, no geometrical significance.
As long as $dr/dR\neq 0$ the simplest choice for $r$ is $R=r$,
i.e.\ Schwarzschild (S) coordinates. In this case
it is common to express $B$ through the ``mass function'' $m$ 
defined by $B=1-2m/r$.

For the {\sl SU(2)\/} Yang-Mills field $W_\mu^a$ we use the
standard minimal spherically symmetric (purely `magnetic') ansatz
\begin{equation}\label{Ans}
W_\mu^a T_a dx^\mu=
  W(R) (T_1 d\theta+T_2\sin\theta d\varphi) + T_3 \cos\theta
d\varphi\;,
\end{equation}
and for the Higgs (triplet) field we assume the form
\begin{equation}\label{AnsH}
\Phi^aT_a=H(R)n^aT_a\;,
\end{equation}
where $T_a$ denote the generators of ${\sl SU(2)\/}$ in the adjoint
representation.
One might also consider other representations for the Higgs , e.g.\ doublet,
but we believe that the behaviour near $r=0$ is the same. 
Plugging these ans\"atze into the EYMH action results in
\begin{equation}\label{Action}
S=-\int dR A
  \Bigl[
  {1\over2}\Bigl(1+B((r')^2
   +{(A^2 B)'\over 2A^2 B}(r^2)')\Bigr)- Br^2V_1-V_2
\Bigr]\;,
\end{equation}
with
\begin{equation}
V_1={(W')^2\over r^2}+{1\over2}(H')^2\;,
\end{equation}
and
\begin{equation}
V_2={(1-W^2)^2\over2r^2}+
{\beta^2r^2\over8}(H^2-\alpha^2)^2+W^2H^2\;.
\end{equation}
Through a suitable rescaling we have achieved that the action depends
only on
the dimensionless parameters $\alpha$ and $\beta$ representing the
mass ratios $\alpha=M_W\sqrt G/g=M_W/g
M_{\rm Pl}$ and $\beta=M_H/M_W$  
($M_H$ and $M_W$ denoting the Higgs resp. gauge  boson mass).

Using S~coordinates the field equations obtained from
(\ref{Action}) are
\begin{subeqnarray}\label{feq}
(BW')'&=&W({W^2-1\over r^2}+H^2)-2rBW'V_1\;,\\
(r^2BH')'&=&(2W^2+{\beta^2r^2\over2}(H^2-\alpha^2))H-2r^3BH'V_1\;,\\
(rB)'&=&1-2r^2BV_1-2V_2\;,\\
A'&=&2rV_1A\;.
\end{subeqnarray}
If $dr/dR=0$ (equator!) S~coordinates become singular
and one has to use a different choice (gauge) for $R$. A convenient
possibility is given by
$B\equiv r^{-2}$ for $B>0$ resp.\ $B\equiv -r^{-2}$ for $B<0$.
We denote this radial coordinate by $\tau$ in order to distinguish it from
the S~coordinate $r$.
With this choice the metric takes the form
\begin{equation}
ds^2=-\frac{A(\tau)}{r^2(\tau)}dt^2+r^2(\tau)(d\tau^2-d\Omega^2)\;.
\end{equation}
We will refer to this system of coordinates as isotropic coordinates.
The equations obtained with isotropic coordinates are 
given in \cite{lav-BLM}.

\section{Singular points}\label{chapsing}

Obviously the field Eqs.~(\ref{feq}) are singular at $r=0$, $r=\infty$ and for
points where $B$ vanishes. 
Note that a generic solution should have five free parameters for the EYMH case
resp.\ three for the EYM case. Here we have ignored the variable $A$, which
decouples (compare Eq.~(\ref{feq}d)).  

At the singular points the generic solution is singular and only a
non-generic subset of solutions stays regular.

In the vicinity of $r=\infty$ one finds a 3-parameter family 
of asymptotically flat solutions 
for the EYMH theory (2-parameter family for the EYM case).

At $r=r_h$ for any given $r_h>0$ one finds a 3-parameter family 
characterized by the value of a gauge and Higgs fields 
at the horizon, $W_h$ and $H_h$ respectively. 

In case of the EYM theory it was shown 
\cite{lav-bholes,lav-existence1,lav-BFM} that for any given
$r_h>0$ there is a discrete set of solutions interpolating 
between the horizon and infinity. 
They can be characterized by an integer $n$, the number of nodes 
of the gauge amplitude $W$.

The EYMH case is more complicated \cite{lav-BFMmonopole,lav-LNW}. 
Let us here only recall that 
the previous research was concentrated on the investigation 
in the interval $[r_h,\infty]$.  

Now we turn to 
the classification of the singular behavior at $r=0$, 
which is of particular relevance for the internal structure of 
black hole solutions.
We have to distinguish two cases, $B>0$ and $B<0$.
\begin{enumerate}
\item[1.]{\bf $B>0$}:\qquad
For black holes this case is only possible, if there is a second, inner
horizon. One finds a 5-parameter, i.e.\ generic, family of solutions.
\begin{subeqnarray}\label{bcrn}
  W(r)&=&W_0+{W_0\over 2(1-W_0^2)}r^2+W_3r^3+O(r^4)\;,\\
  H(r)&=&H_0+H_1r+O(r^2)\;,\\
  B(r)&=&{(W_0^2-1)^2\over r^2}-{2M_0\over r}+O(1)\;,
\end{subeqnarray}
where  $W_0^2\neq 1$.

According to the asymptotics of $B(r)$ we may call the singular
behavior to be of RN-type.
The special case $W_0^2=1, M_0<0$ leads to S-type behavior with a
naked singularity.
On the other hand $W_0^2=1, H_0=0, M_0=0$ gives regular solutions.

\item[2.] {\bf $B<0$}:\qquad
This case is more involved, with two disjoint families of singular
solutions.

\item[2.1]
There is a 3-parameter family of solutions with a S-type singularity,
characterized by the asymptotics
\begin{subeqnarray}\label{bcss}
  W(r)&=&1+W_2r^2+O(r^3)\;,\\
  H(r)&=&H_0+O(r)\;,\\
  B(r)&=&-{2M_0\over r}+O(1)\;.
\end{subeqnarray}
where $M_0>0$. 

Obviously the condition $B<0$ ($M_0>0$) prevents the existence of
regular solutions in this case.

\item[2.2 ]
There is an additional 2-parameter family of solutions with a pseudo-RN
singularity (pseudo because $B<0$).
\begin{subeqnarray}\label{bcqrn}
  W(r)&=&W_0\pm r+O(r^2)\;,\\
  H(r)&=&H_0+O(r^2)\;,\\
  B(r)&=&-{(W_0^2-1)^2\over r^2}\pm{4W_0(1-W_0^2)\over r}+O(1)\;.
\end{subeqnarray}
with $W_0^2\neq 1$. 
The eigenvalues of the linearized equations are
$\lambda_{1,2}=-1/2(3\pm i\sqrt 15)$ and $\lambda_3=-1$.
This is a repulsive focal point which will turn out to be important for the
cyclic behavior in the EYM case.
\end{enumerate}

Note that the corresponding
Taylor series for the singular points in case of the EYM theory were 
listed (with minor mistakes) in \cite{lav-DGZ1}, 
whereas in our work \cite{lav-BLM} the more general EYMH theory was studied
and local existence proof was given. In other words we have  
shown that the expressions above are in fact the beginning 
of a {\bf convergent} Taylor series for $W,H$ and $r^2B$.

In the case $B<0$ we obtained no singular class that has enough parameters
(three for EYM and five for EYMH) to describe the generic behavior. Since,
also the appearance of a second, inner horizon is a
non-generic phenomenon, one may wonder, what the generic behavior inside
the horizon near $r=0$ looks like. 
This  situation is shown schematically on the 
Fig.~\ref{branches} for the case of the EYM theory.

\section{Numerical results}\label{chapnum}

In order to investigate the generic behavior of non-Abelian black
holes inside the event horizon, we integrate the field Eqs.~(\ref{feq}) 
\footnote{
At least part of the solutions possess an equator, i.e.\ a 
local maximum of $r$. For those the use of S~coordinates is excluded
and we integrated the equations in the isotropic coordinates.}
from the horizon assuming $B<0$, 
ignoring the constraints on the initial data at the
horizon required for asymptotic flatness.

Using the Killing time $t$ the
horizon is a singular point of the equations. Consequently one has to
desingularize the equations in order to be able to start the integration
right there. How this can be done, was described 
in \cite{lav-BLM,lav-BFMmonopole}.

As one performs the numerical integration one quickly runs into 
problems due to the occurrence of a quasi-singularity, initiated
by a sudden steep raise of $W'$ and subsequent exponential growth
of $B$ resp.\ $m$ (compare Figs.~\ref{expl2}, and~\ref{explh}
for some examples).
This inflationary behaviour of the mass function
is similar to the one observed for perturbations of the 
Abelian black hole solution at the Cauchy horizon 
\cite{lav-PI,lav-Ori1,lav-Page,lav-Ori2}. 
While this fast growth continues
indefinitely for the EYMH system, it comes to a stop without the Higgs
field. The mass function reaches a plateau and stays constant for a
while until it starts to decrease again. When $B$ has become small
enough, i.e.\ the solution comes close to an inner horizon, the same
inflationary process repeats itself. Generically this second
``explosion'' is so violent (we will give estimates on the increase
of $m$ in chapter~\ref{chapqual}) that the numerical integration
procedure breaks down.

Besides these generic solutions there are certain families of special
solutions obtained through suitable fine-tuning of the initial data at
the horizon.  There are two classes of such special solutions.  The
first class are black holes with a second, inner horizon, the second are
solutions with one of the singular behaviours at $r=0$ for $B<0$
described in chapter~\ref{chapsing}.  The numerical construction of such
solutions is complicated by the fact that both boundary points are
singular points of the equations.  The strategies employed to solve such
problems are well described in the paper on gravitating monopoles
\cite{lav-BFMmonopole}.  
Actually, in order to control the numerical uncertainties
we used two different methods, which may be called ``matching'' and
``shooting and aiming''.  For matching we integrate independently from
both boundary points with regular initial data, tuning these data at
both ends until the two branches of the solution match.  For shooting
and aiming we integrate only from one end and try to suppress the
singular part of the solution at the other end by suitably tuning the
initial data at the starting point.

Our results concerning special solutions are shown in Fig.~\ref{curves}
and discussed in \cite{lav-BLM}.
As already said, the first class of special solutions
consists of black holes with a second, inner horizon; 
we call them non-Abelian RN-type (NARN) solutions.
We have determined two such 1-parameter families for the EYM system,
shown in Fig.~\ref{curves}.
As may be inferred from Fig.~\ref{curves}, the (dotted) curve~2
corresponding to one such family intersects all (solid) curves
describing asymptotically flat solutions except the one for $n=1$.
Curve continues straight through the parabola
$r_h=1-W_h^2$ and runs all the way to $r_h=W_h=0$.
The branch to the left of the parabola cannot be obtained using 
S~coordinates since solutions develop local maximum of $r$
between the horizons.

Our second NARN family (curve~5 of Fig.~\ref{curves}) stays
completely to the left of the parabola and ends at $r_h\approx0.9$ close
to the curve~3, whose significance will be explained below.

The second class are solutions without a second horizon
(i.e.\ $B$ stays negative) approaching the center $r=0$ with one of the
two singular behaviours described in chapter~\ref{chapsing},
i.e.\ those with a S-type singularity resp.\ with a pseudo-RN-type
singularity; we denote them NAS resp.\ NAPRN solutions.
We have determined several NAS families
represented by the dashed-dotted curves of Fig.~\ref{curves}.
The curve~1 staying to the right of the parabola coincides with the
corresponding one found in \cite{lav-DGZ1}, whereas the others, 
staying essentially to the left of the parabola are new \cite{lav-BLM}. 
The basic NAS curve~1 intersects (once) only $n=1$ (solid) curve for 
asymptotically flat black holes. 
As will be explained in chapter~\ref{chapqual}, 
the two NAS curves~6 and~7 accompanying the (dotted)
NARN curve~5 are expected to merge with the NAS curve~3 close to
$r_h=0.9$. Some of the NAS curves (e.g., 3 and~4) are expected to extend
indefinitely to the right, but numerical difficulties (too violent
``explosions'') prevented us from continuing them further to larger values
of $r_h$. They will intersect the (solid) curves for asymptotically flat
solutions with $n=2,3,\ldots$ zeros of $W$ and therefore yield additional
asymptotically flat NAS black holes.

Asymptotically (for big $r_h$) the basic NARN curve~2 approaches the 
basic NAS curve~1. 
Why this happens can be ``understood'' from Fig.~\ref{narn0}.

Finally there are the NAPRN solutions, which constitute a discrete set
according to the number of available free parameters at $r=0$.
We found  several such solutions \cite{lav-BLM}.
Few NAPRN solutions are shown in 
Fig.~\ref{naprn12} and Fig.~\ref{naprn34}.
Only one of them has no maximum of $r$ and was found in 
\cite{lav-DGZ1} as well.
Let us stress that although the NAPRN solutions do not correspond to 
asymptotically flat black holes they play essential role in the 
explanation on the cyclic behaviour of a generic solution
in the EYM case (compare discussion in the next chapter~\ref{chapqual}). 

\section{Qualitative Discussion}\label{chapqual}

We shall now give a qualitative picture of the solutions and try to
explain our numerical results. Let us briefly summarize 
what can be done (and what in fact has been done \cite{lav-DGZ1,lav-BLM})

-- get a qualitative understanding of the solutions \cite{lav-DGZ1,lav-BLM}

-- obtain a plateau -- to -- plateau formula  in the EYM case
relating quantities at one plateau (before ``explosion'') to those 
on the next plateau (after ``explosion'') \cite{lav-BLM} 

-- describe a simplified dynamical system, 
which reflects the main properties of the generic 
solutions \cite{lav-DGZ1,lav-BLM} 

Since the generic behaviour of the
solutions is rather different in the cases with and without Higgs field,
we shall treat the two cases separately. Let us first concentrate
on the case without Higgs field.

\subsection{EYM theory}

For a ``naive'' understanding of the cyclic behaviour one can use a
mechanical analogy. Introducing the ``time'' variable 
$\sigma=-\ln(r)$ the EYM equations can be written in the form  
\begin{subeqnarray}\label{manalogy}
\ddot{W}&=&{{W(W^2-1)}\over B} -\left[2+{1\over B}
\left({{(W^2-1)^2}\over {r^2}} -1\right)\right]\dot{W}\;,\\
\dot{B}&=&\left({{(W^2-1)^2}\over {r^2}}-1\right)
+\left(1+{{2\dot{W}^2}\over {r^2}}\right)B\;,
\end{subeqnarray}
where $\dot{}\equiv d/d\sigma$.
The first equation Eq.~(\ref{manalogy}a)) 
resembles the motion of a ficticious particle
in a potential with velocity dependent friction.

Note that the sign of the friction coefficient (term in square brackets 
in  the Eq.~(\ref{manalogy}a))  can be positive as well as negative. 
Close to the horizon ${{(W^2-1)^2}/{r^2}} -1<0$ and friction coefficient 
is positive, corresponding to a deceleration of the ``particle''.
As the time $\sigma$ increases ($r$ decreases) this term changes sign and 
the friction 
turns into anti-friction. The particle starts to accelerate quickly.
This leads to a domination of the second term (kinetic energy) in 
Eq.~(\ref{manalogy}b), which in turn leads to a fast growth 
of the function $B$ (respectively $m$). But growth of  $B$ stops the   
anti-friction in Eq.~(\ref{manalogy}a) and the particle is again in the 
slow roll regime until the next ``explosion''. 

For more detailed discussion of the generic behaviour
we introduce the notation $\bu\equiv BW'$ and $\bb\equiv rB$ and use 
again $\sigma\equiv -\ln(r)$ as a radial coordinate \cite{lav-BLM}. 
Note  that $\bb\approx -2m$ for small $r$.
With these variables the field Eqs.~(\ref{feq})

\begin{subeqnarray}\label{bareq}
\dot{W}&=&-r^2{\bu\over\bb}\;,\\
\dot{\bu}&=&-W{W^2-1\over r}+
2r^2{\bu^3\over\bb^2}\;,\\
\dot{\bb}&=&r\biggl({(1-W^2)^2\over r^2}-1\biggr)+2r^2{\bu^2\over\bb}\;.
\end{subeqnarray}
Close to the horizon the first term in the equation for $\bb$
dominates (since $\bu$ vanishes at $r=r_h$) and thus $\bb$
becomes negative. Provided $W^2$ does not tend to $1$,
this term will, however, change sign as $r$ decreases 
and $\bb$ will turn back to zero.
Assuming further that $\bu$ does not tend to zero simultaneously,
the second term in the equation for $\bu$ will grow very rapidly
as $\bb$ approaches zero, leading to a rapid increase of $\bu$.
This in turn induces a rapid growth of $\bb$ (compare Fig.~\ref{expl2}).
Once the second terms in Eqs.~(\ref{bareq}b,c)
dominate one gets
${(\bu/\bb)}\dot{}\approx 0$ and thus $\bu/\bb=W'/r$ tends to a constant
$c$.
As long as $(rc)^2$ is sizable $\bu$ and $\bb$ increase exponentially,
giving rise to the phenomenon of mass inflation.
Eventually this growth comes to a stop when $(rc)^2$ has
become small enough. Then $\bu$ and $\bb$ stay constant until
the first terms in Eqs.~(\ref{bareq}b,c) become sizable again.
As before $\bb$ tends to zero inducing another ``explosion'' resp.\
cycle of mass inflation (compare Fig.~\ref{expl2}).

In the discussion above we made two provisions -- that
$W^2$ stays away from $1$ and that $\bu$ does not tend zo zero
simultaneously with $\bb$. If the first condition is violated, i.e.\
$W^2\to 1$ we get a NAS solution. If on the other hand $\bu$ and $\bb$
develop a common zero we get a NARN solution, i.e.\ a solution
with a second horizon. Both these phenomena can occur after any finite
number of cycles, giving rise to several NAS resp.\ NARN curves as in
Fig.~\ref{curves}.
Generically $W$ changes very little during an inflationary cycle,
with the exception of solutions that come very close to a second
horizon, i.e.\ close to a NARN solution. In this case $W$ may change by
any amount, depending on how small $\bu$ becomes at the start of the
explosion. By suitably fine-tuning the initial data at the horizon one
can then obtain new NAS solutions with $W\to\pm1$ or a new NARN solution. In
this way each NARN solution is the `parent' of two NAS and one NARN
solution. This schematically shown on the Fig.~\ref{children}. 
Fig.~\ref{curves} shows two such generations: the NARN solutions
labelled~2 have the NAS children~3 and~4 and the NARN child~5; the curves
labelled~6 and~7 are the NAS children of~5.
Whenever the value of $W$ at the second horizon of a NARN solution approaches
$\pm 1$ this NARN curve and its NAS children merge with the corresponding
sibling NAS curve having one cycle less. This hierarchy of special solutions
gives rise to a kind of chaotic structure in this region of ``phase space''.

Neglecting irrelevant terms one can integrate 
Eqs.~(\ref{bareq}) and obtain a plateau -- to -- plateau  relation
\cite{lav-BLM}, which connects the quantities $W_0,\bu_0$ and $\bb_0$ 
before an explosion with $W_1,\bu_1,\bb_1$ after it
\footnote{Note that the similar relations are obtained 
recently in \cite{lav-GDZ2}.}

\begin{equation}\label{map}
\bu_1=\bu_0e^{(cr_0)^2}\;,\quad \bb_1={\bu_0\over c}e^{(cr_0)^2}\;,
\quad W_1=W_0-{c\over2}r_0^2\;,
\end{equation}
with
\begin{equation}
r_0=-{(W_0^2-1)^2\over\bb_0}\;,\quad
c={(W_0^2-1)^2\over2\bu_0r_0^3}\;.
\end{equation}
It is instructive to illustrate these relations on an example.
We take the fundamental black hole solution with $r_h=1$ and
$W_h=0.6322$ shown in Fig.~1 in \cite{lav-BLM}.
For the first explosion one finds the parameters
$r_0\approx 2.7\cdot 10^{-4}$
and $c\approx 1.1\cdot 10^5$
yielding $cr_0\approx 30$ and thus $\bb_1\sim e^{900}$ and
$W_1-W_0\approx 4\cdot 10^{-3}$.
The subsequent explosion will then take place at the fantastically small
value $r_0\sim e^{-900}\approx 10^{-330}$.

Since the change of $W$ in one inflationary cycle has an extra factor
$r_0$ the function $W$ stays practically constant. If we furthermore concentrate
on cases, where the first term in Eq.~(\ref{bareq}b) can be neglected
we may use the simplified system \cite{lav-DGZ1,lav-BLM}
\begin{equation}\label{simpeq}
\dot W=0\;,\quad
\dot\bu=2r^2{\bu^3\over\bb^2}\;,\quad
\dot\bb={(1-W^2)^2\over r}+2r^2{\bu^2\over\bb}\;.
\end{equation}
Introducing the variables $x\equiv r\bu/\bb=W'$ and
$y\equiv -(1-W^2)^2/r\bb$ one obtains the autonomous system
\begin{equation}\label{autoeq}
\dot W=0\;,\quad
\dot x=(y-1)x\;,\quad
\dot y=y(y+1-2x^2)\;.
\end{equation}
Since the first of these equations may be ignored, we can concentrate
on the $x,y$ part. As usual for 2-dimensional dynamical systems the
global behavior of the solutions can be analyzed determining its fixed
points. Since the ``large time'' behavior $\sigma\to\infty$ corresponds to
the limit $r\to 0$ these fixed points are related to the singular solutions
at $r=0$ discussed in chapter~\ref{chapsing}.
There are essentially three different fixed points.
\begin{enumerate}
\item[1.]
For $y<0$ there is the fixed point $x=0$, $y=-1$ giving the RN type
singularity. Its eigenvalues are $-1$ and $-2$, hence it
acts as an attracting center for $\sigma\to\infty$.
\item[2.]
Then there is the point $x=y=0$, a saddle with eigenvalues $\pm1$.
\item[3.]
In addition there are the points $x=\pm1$, $y=1$ with the eigenvalues
$1/2(1\pm i\sqrt15)$, related to the pseudo-RN type singularity.
This fixed point acts as a repulsive focal point, from which the
trajectories spiral outwards. Since solutions of the approximate system
given by the Eqs.~(\ref{autoeq}) cannot cross the coordinate axes,
solutions in the quadrants $y>0$, $x>0$ resp.\ $x<0$ stay there
performing larger and larger turns around the focal point coming closer
and closer to the saddle point $x=y=0$ without ever meeting it.
As observed in \cite{lav-DGZ1} this nicely explains the cyclic
inflationary behaviour of the solutions in the generic case.
\end{enumerate}

It is interesting to note that the similar results were obtained 
in the Abelian case \cite{lav-Page,lav-Ori2} in the homogeneous 
mass inflation model
\footnote{We are thankful to A.Ori for bringing the 
ref.~\cite{lav-Page} to our attention and for communicating his 
unpublished results \cite{lav-Ori2}.}.

\subsection{EYMH theory}

Finally let us discuss the black holes with Higgs field.
Apart from the generic solutions there are the special ones approaching
$r=0$ with a singular behaviour described in chapter~\ref{chapsing}.
On the other hand, the generic behaviour is much simpler than in the
previously discussed situation without Higgs field.
An easy way to understand this difference is to derive the analogue of
the simplified system Eqs.~(\ref{autoeq}). Introducing the additional
variable $z\equiv -\dot{H}$ and ignoring again irrelevant
terms one finds \cite{lav-BLM}
\begin{subeqnarray}\label{autoheq}
\dot{W}&=&0\;,\quad\dot{H}=-z\;,\\
\dot{x}&=&(y-1)x\;,\quad\dot{z}=yz\;\\
\dot{y}&=&y(y+1-2x^2-z^2)\;.
\end{subeqnarray}
Leaving aside the decoupled equations for $W$ and $H$ one may study the
fixed points of the $(x,y,z)$ system.  For $z=0$ one clearly finds the
previous fixed points of the $(x,y)$ system.  However, for $z\neq0$ the
focal point disappears and the only fixed point for $y\ge0$ is $x=y=0$,
$z=z_0$ with some constant $z_0$.  For $z_0^2<1$ this point is a saddle
with one unstable mode, whereas for $z_0^2>1$ it is a stable attractor.
The latter describes the simple inflationary behavior described in
chapter~\ref{chapnum} and shown in the left part of Fig.~\ref{explh}.
Solutions approaching a fixed point with $z_0^2<1$ eventually run away
from it again and ultimately tend to one with $z_0^2>1$ as shown in the
right part of Fig.~\ref{explh}.  

In the limit the equations can be trivially integrated with the result 
\begin{equation}\label{hfpoint}
y=y_{0} e^{(1-z^{2}_{0})\sigma}=y_{0} r^{z^{2}_{0}-1}.  
\end{equation}
Thus the mass function grows exponentially in ``time'' $\sigma$
or as a power in terms of $r$
\begin{equation}\label{mass}
m=m_{0} e^{z^{2}_{0}\sigma}={m_{0}\over {r^{z^{2}_{0}}}}  
\end{equation}
in perfect agreement with our numerical results Fig.~\ref{explh}. 
Note that the fixed point Eq.~(\ref{hfpoint}) differs from the 
ones listed in the  chapter~\ref{chapsing} since the nature (strength) 
of the singularity depends on the solution itself.
This fixed point was ``rediscovered'' in \cite{lav-GD}.

Note that the inclusion of a scalar field 
in the homogeneous mass inflation model \cite{lav-Ori2} 
has the same effect as the addition of the Higgs field to the EYM theory, 
namely that the mass inflation cycles disapear.

\section{Concluding remarks}\label{chapconc}

To summarize the interior geometry of non-Abelian black holes 
exhibits a very interesting and complicated structure.
Besides the generic solutions there are special NAS, NARN and NAPRN solutions, which can be 
obtained by a fine tuning initial data at the horizon. 
The main conclusion is that
no inner (Cauchy) horizon is formed inside non--Abelian black holes
in generic case, instead one obtains a kind of mass inflation.
Without a Higgs field, i.e.\ for the EYM theory, this mass inflation 
repeats itself in cycles of ever more violent growth.

A natural question to ask is 
what might be potential outcome of this investigation.
A short answer would be

-- illustration of singularity theorems \cite{lav-HE}

-- possible cosmological applications (see e.g. \cite{lav-GD})

-- interesting laboratory for non-perturbative 
study of (homogeneous) mass inflation phenomenon

Naturally one should bare in mind limitations
which come from quantum corrections and instability.
Our consideration was purely classical, but 
when $r$ tends to zero the curvature diverges and quantum 
corrections will become important.
As it was discussed in the Introduction the EYM and most 
of the EYMH black holes are classically unstable.
What will be a fate of a non-static perturbations inside 
non-Abelian black holes is an interesting open question.
Definitely these subjects require further study.
  
\section{Acknowledgments}

G.L. is grateful to the organizers for the invitation 
to Haifa and kind hospitality extended during the conference.

The work of G.L. was supported in part by the Tomalla Foundation and by
the Swiss National Science Foundation.

\newpage

\begin{figure}
\hbox to\hsize{\hss
   \epsfig{file=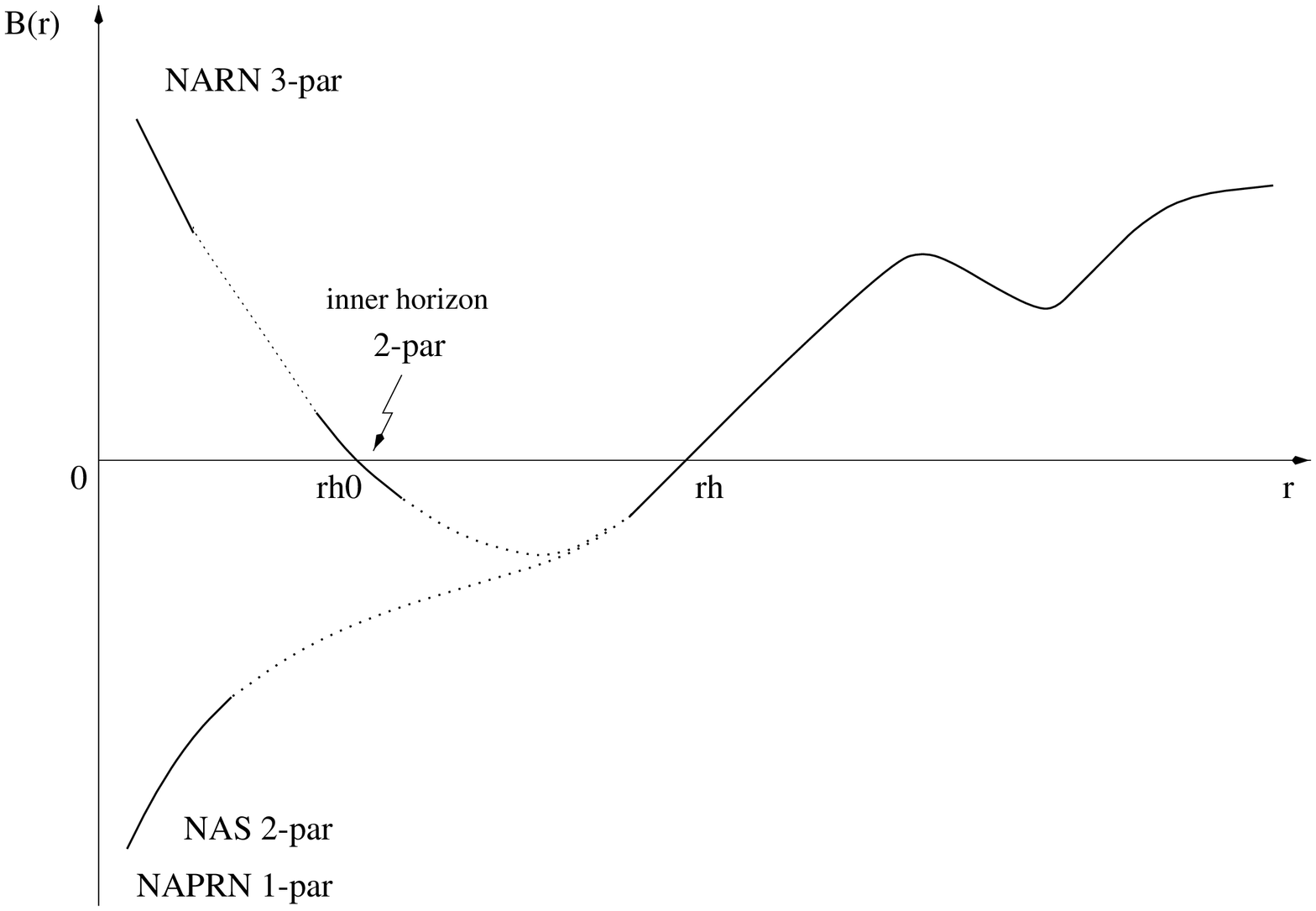,width=\hsize,%
   bbllx=1.3cm,bblly=0.0cm,bburx=20.0cm,bbury=19.8cm}%
  }
\labelcaption{branches}{
Schematic view of the different special solutions in the EYM theory.}
\end{figure}

\newpage

\begin{figure}
\hbox to\hsize{\hss
   \epsfig{file=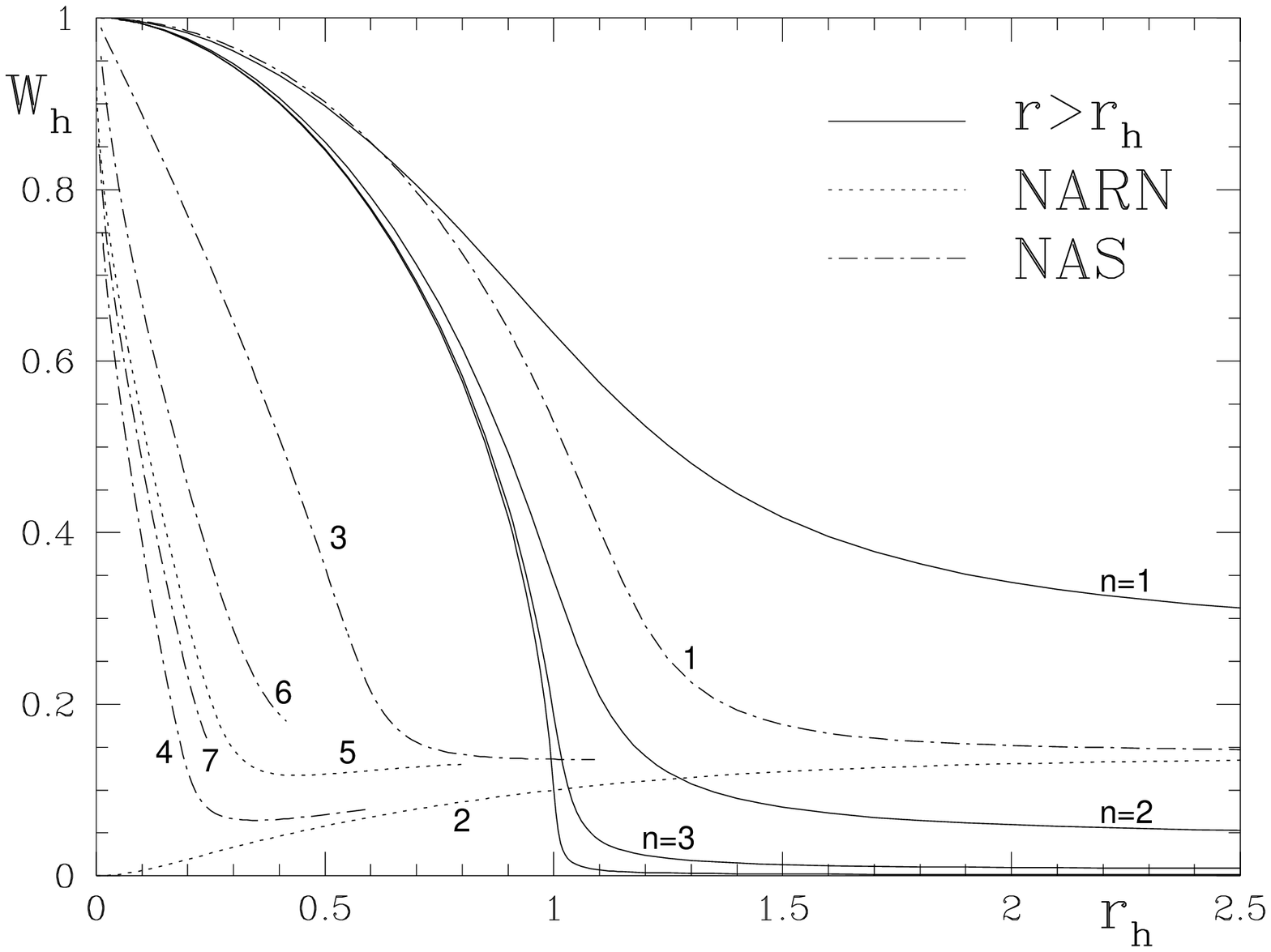,width=\hsize,%
   bbllx=1.3cm,bblly=6.1cm,bburx=20.0cm,bbury=19.8cm}%
  }
\labelcaption{curves}{
Initial data for special solutions. The solid curves represent
asymptotically flat solutions with $n$ zeros of $W$. The other curves
represent various NARN and NAS families.}
\end{figure}

\newpage

\begin{figure}
\hbox to\hsize{\hss
   \epsfig{file=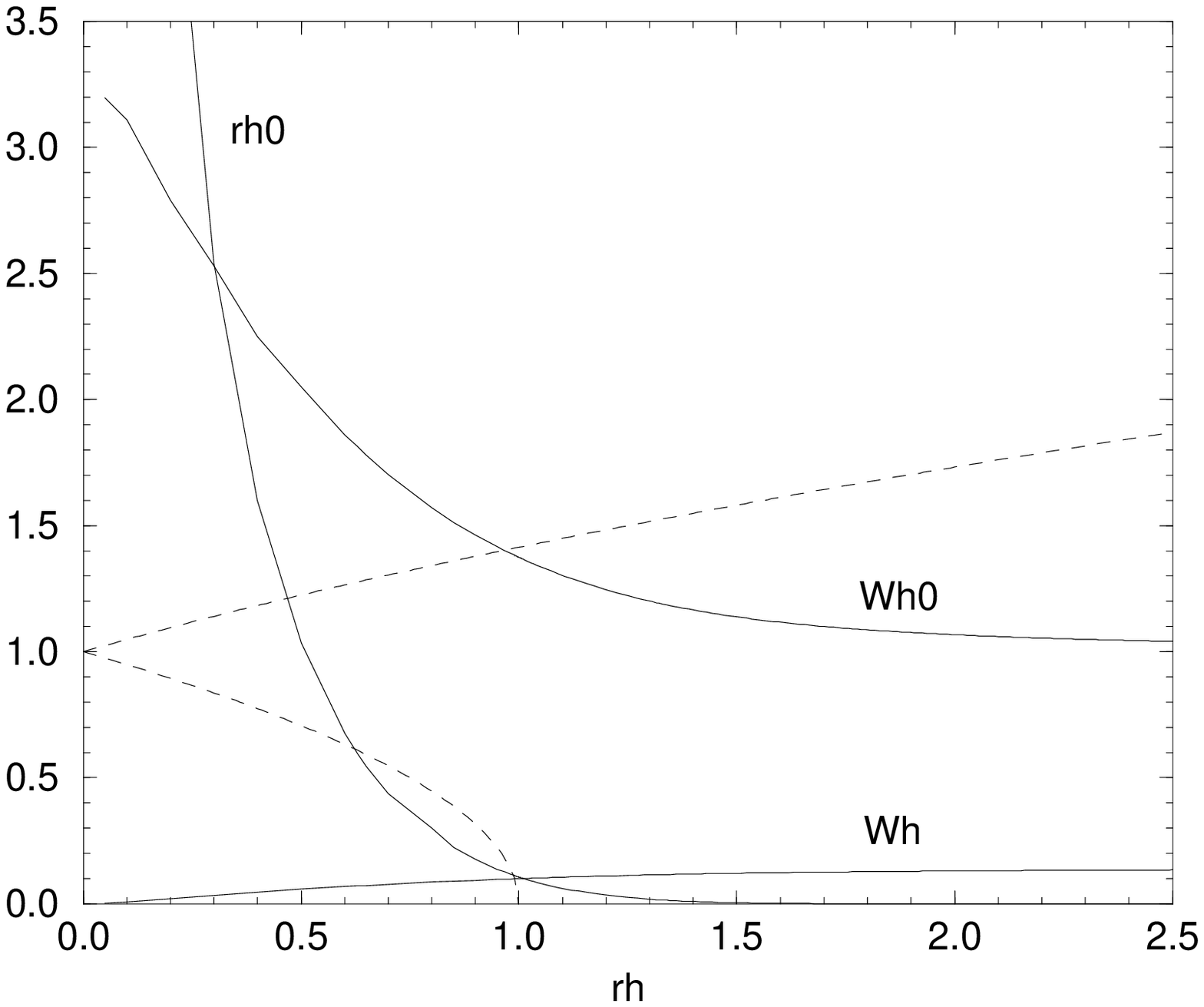,width=\hsize,%
   bbllx=1.3cm,bblly=0.0cm,bburx=20.0cm,bbury=19.8cm}%
  }
\labelcaption{narn0}{
Parameters of the basic (lowest) NARN solutions.
One clearly sees that with increasing $r_h$ the value of the
second (inner) horizon $r_{h0}\to 0$ and value of the gauge 
field there $W_{h0}\to 1$. This values correspond to NAS solutions.}
\end{figure}

\newpage

\begin{figure}
\hbox to\hsize{
  \epsfig{file=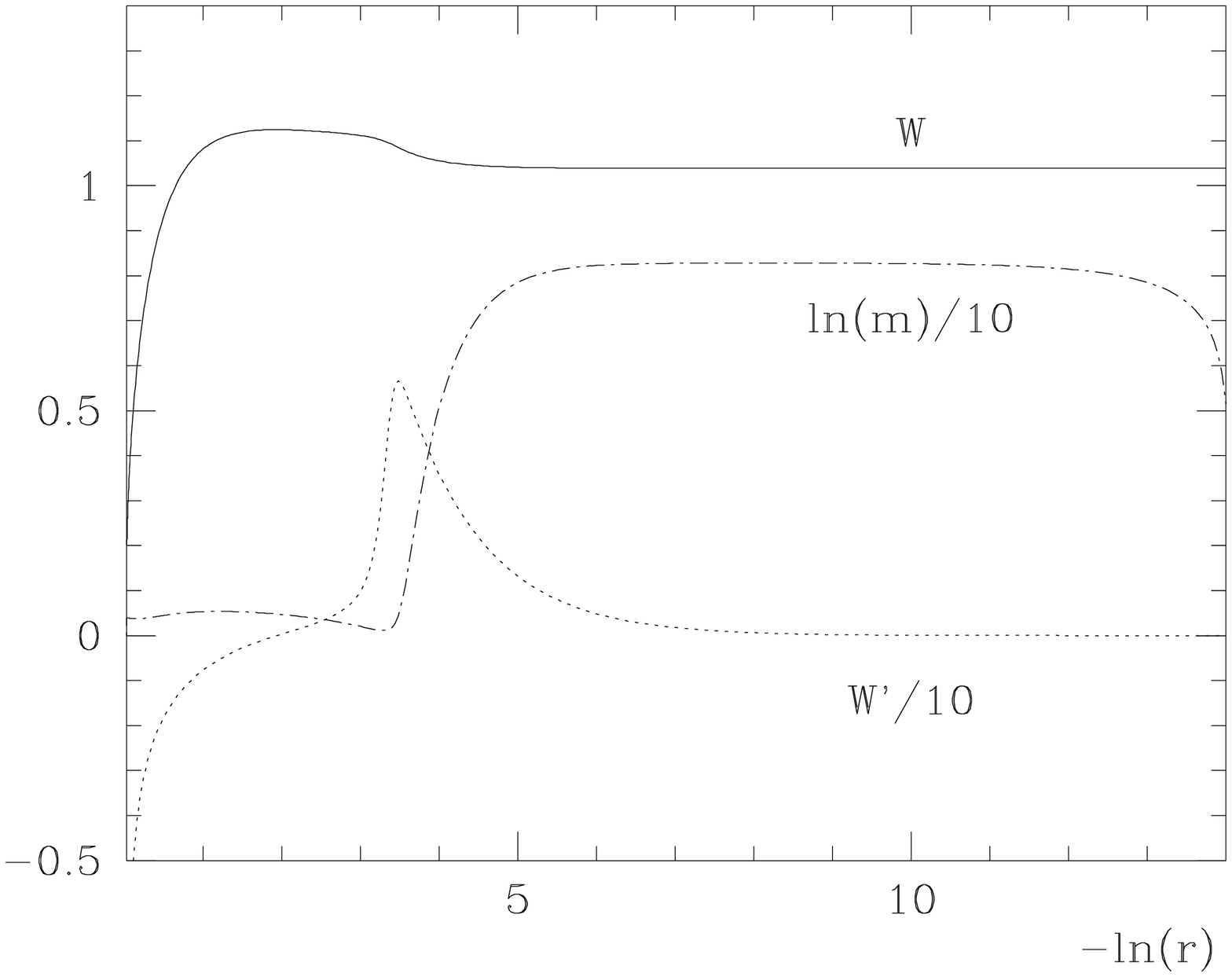,width=0.48\hsize,%
      bbllx=2.3cm,bblly=5.9cm,bburx=20.0cm,bbury=20.0cm}\hss
  \epsfig{file=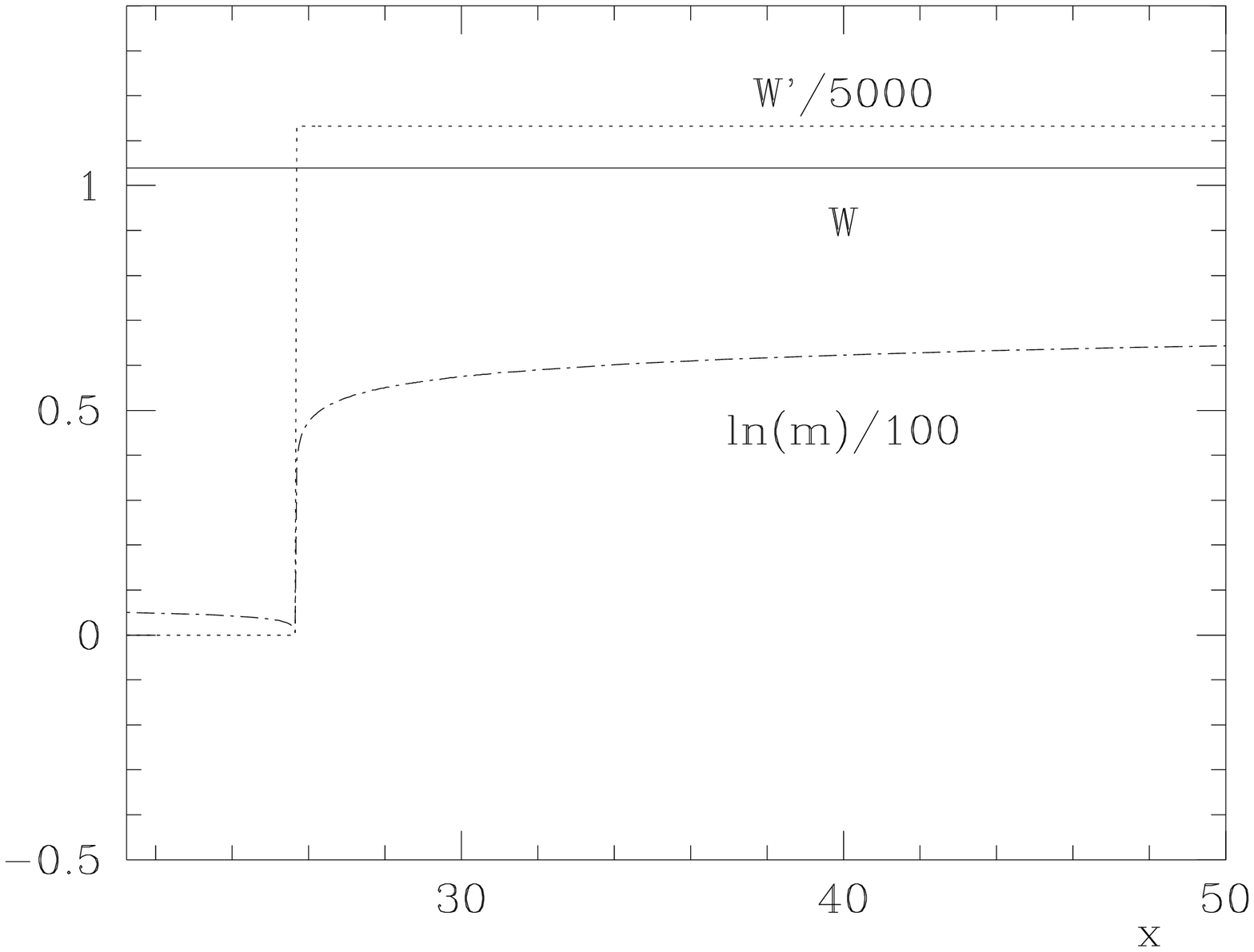,width=0.48\hsize,%
      bbllx=1.8cm,bblly=5.5cm,bburx=20.0cm,bbury=20.0cm}
  }
\labelcaption{expl2}{
First two cycles of the solution with  $r_h$=0.97 and $W_h=0.2$.  For
the second cycle a suitably stretched coordinate $x$ is used.}
\end{figure}
\begin{figure}
\hbox to\hsize{
  \epsfig{file=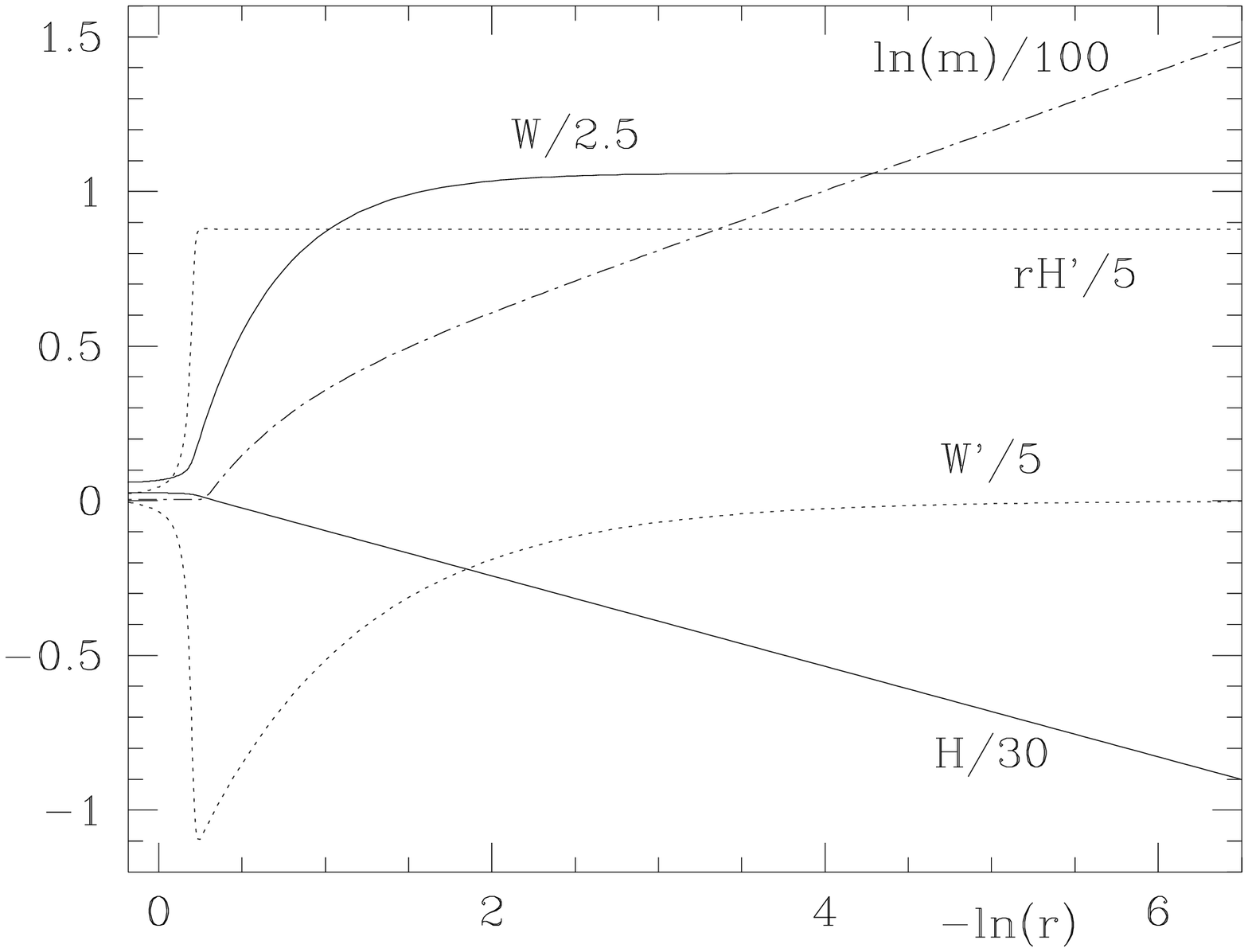,width=0.48\hsize,%
      bbllx=2.3cm,bblly=5.9cm,bburx=20.0cm,bbury=20.0cm}\hss
  \epsfig{file=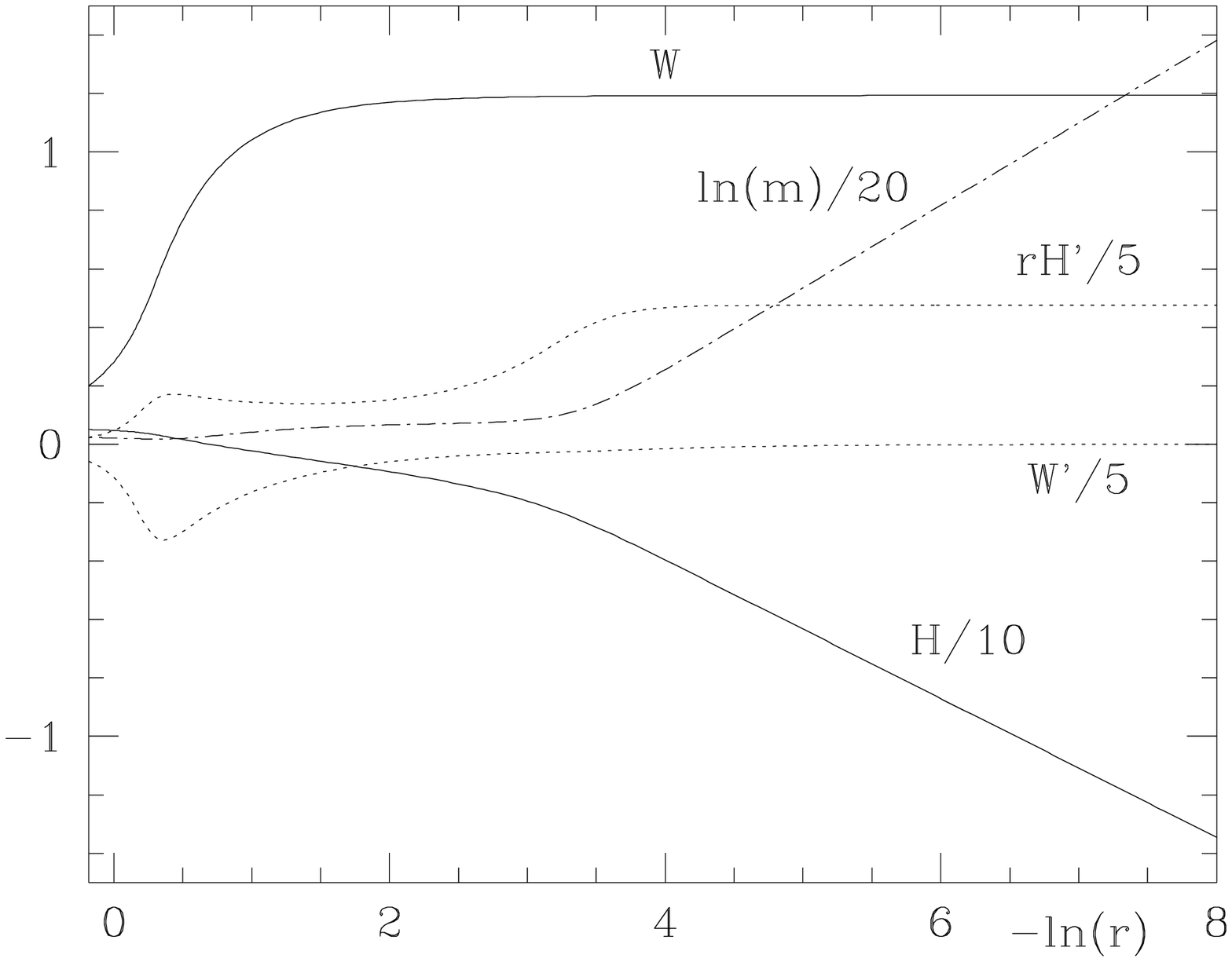,width=0.48\hsize,%
      bbllx=1.8cm,bblly=5.5cm,bburx=20.0cm,bbury=20.0cm}
  }
\labelcaption{explh}{
Two characteristic types of inflationary solutions with Higgs fields.
Note that in both cases asymptotically $\ln(m)$ is linear in $\ln(r)$.} 
\end{figure}

\newpage

\begin{figure}
\hbox to\hsize{
  \epsfig{file=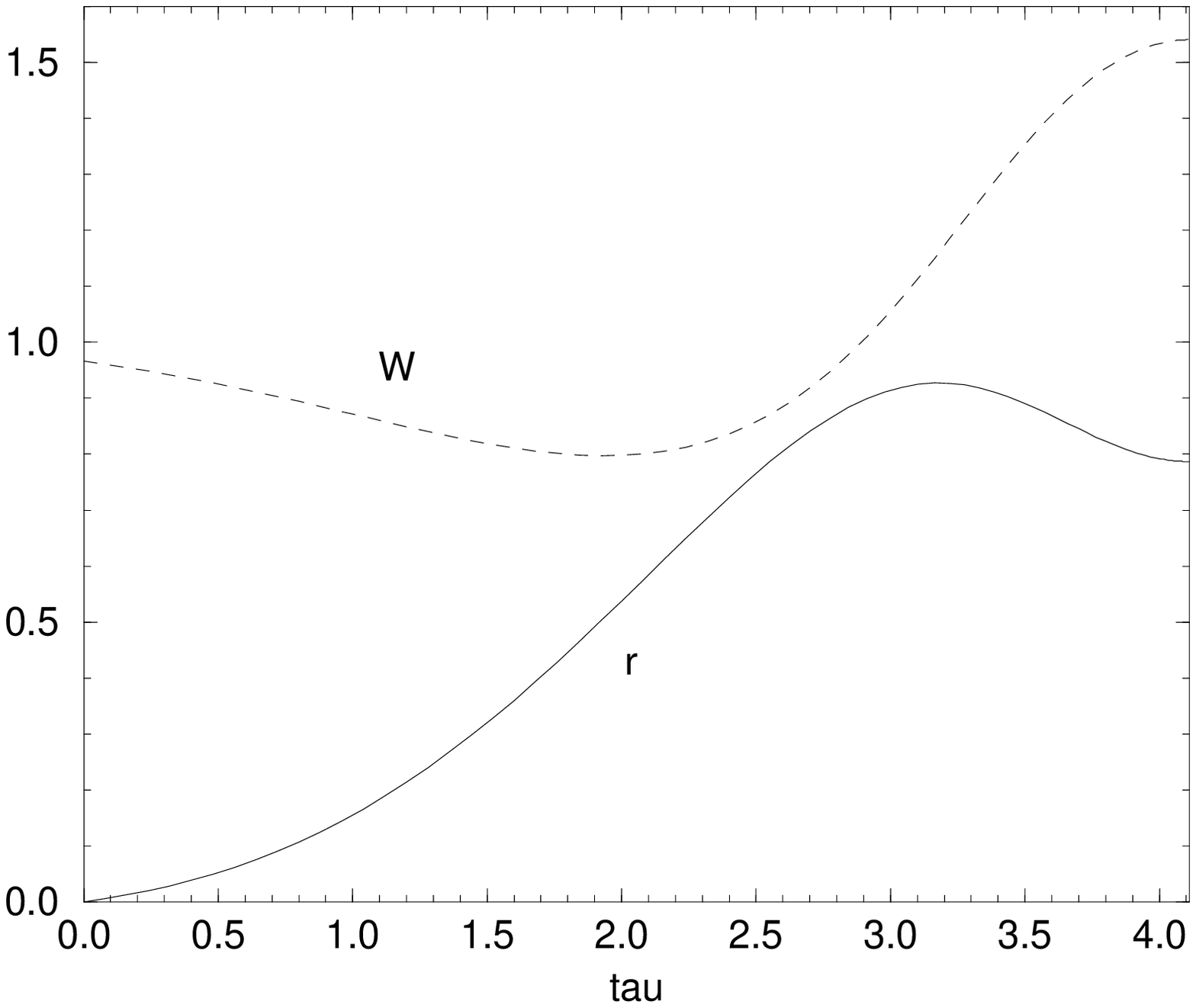,width=0.48\hsize,%
      bbllx=2.3cm,bblly=1.5cm,bburx=20.0cm,bbury=20.0cm}\hss
  \epsfig{file=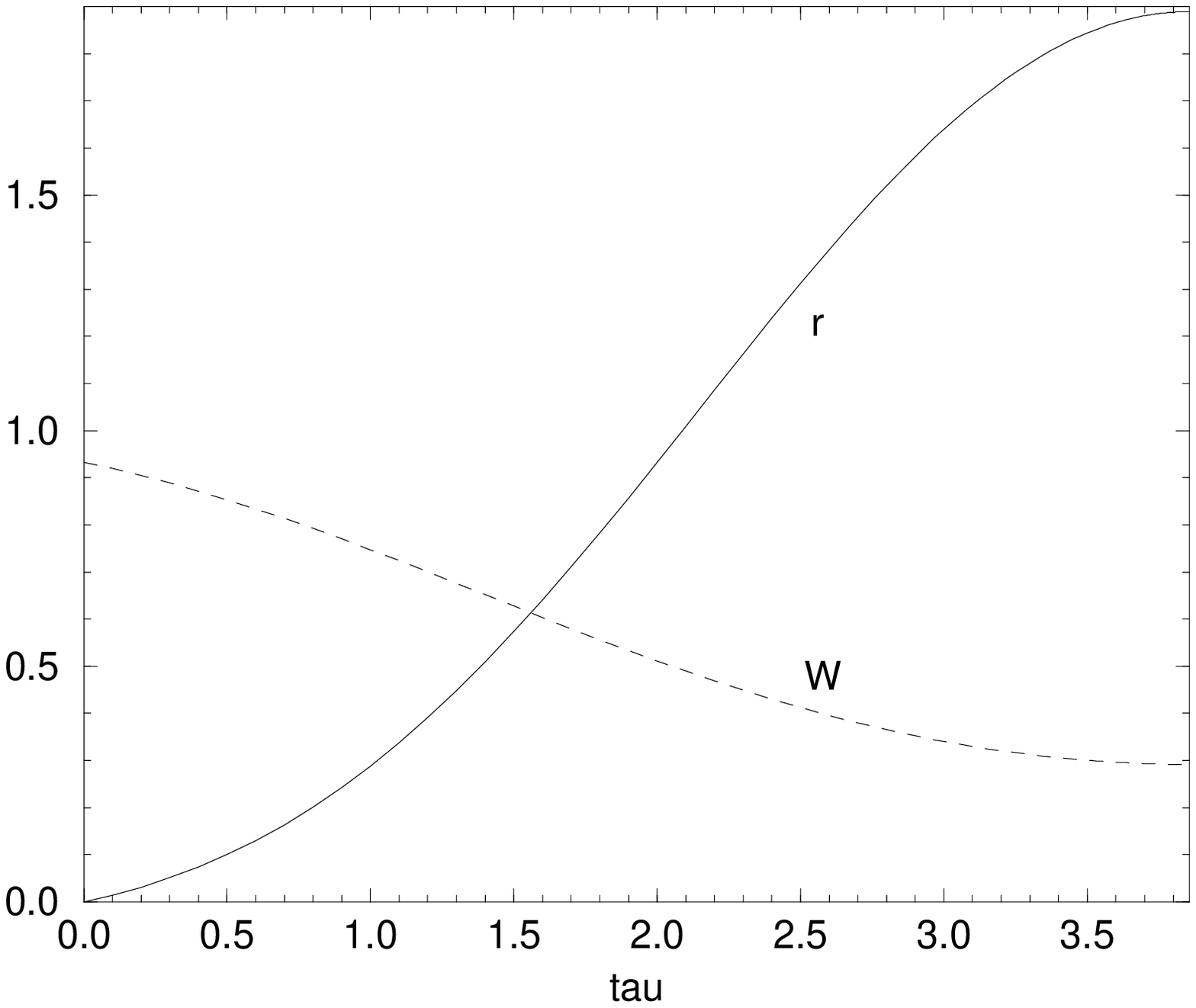,width=0.48\hsize,%
      bbllx=1.8cm,bblly=1.5cm,bburx=20.0cm,bbury=20.0cm}
  }
\labelcaption{naprn12}{
Two different NAPRN solutions with no zero of $W$ between $r=0$ and $r=r_h$.}
\end{figure}
\begin{figure}
\hbox to\hsize{
  \epsfig{file=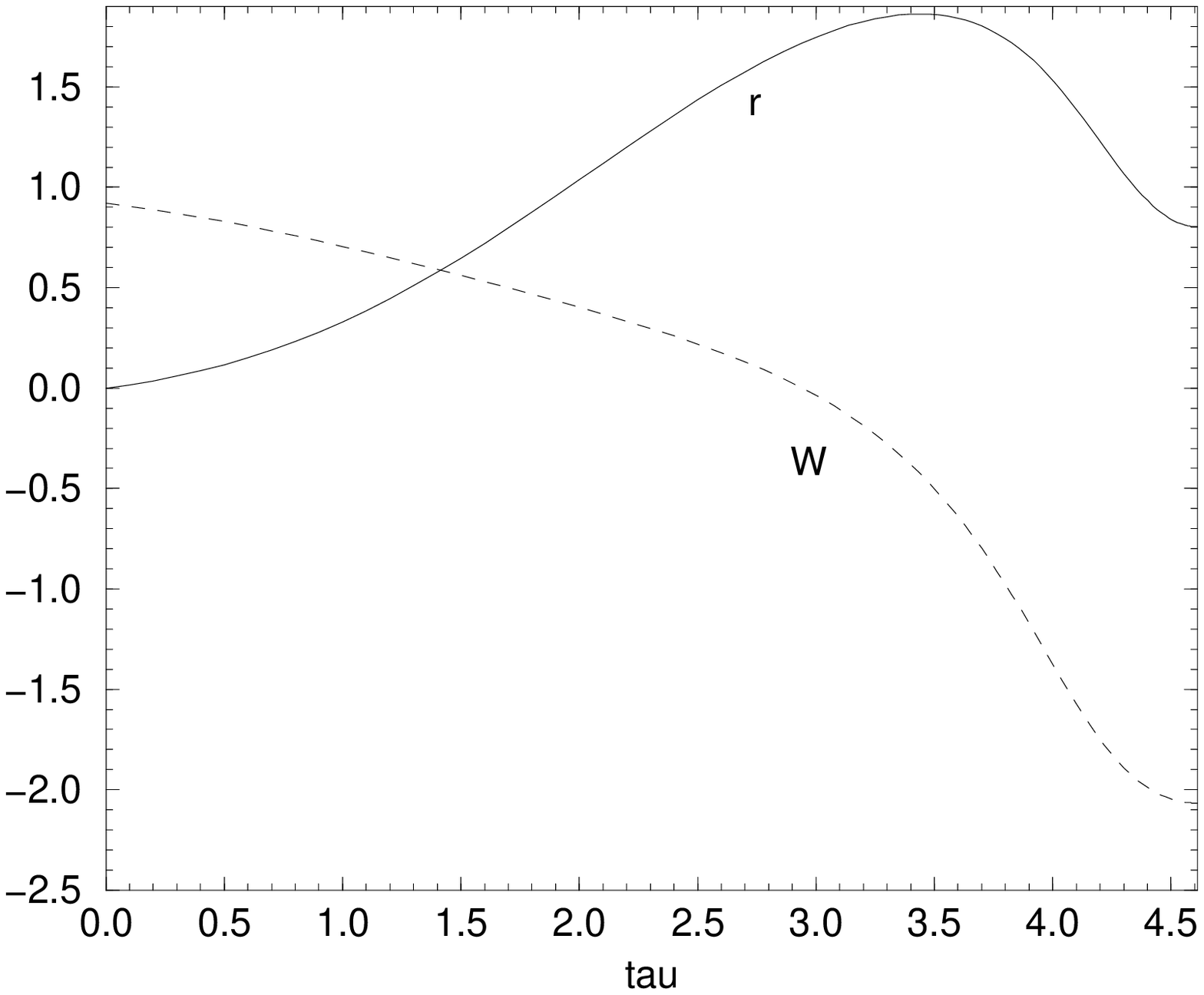,width=0.48\hsize,%
      bbllx=2.3cm,bblly=1.5cm,bburx=20.0cm,bbury=20.0cm}\hss
  \epsfig{file=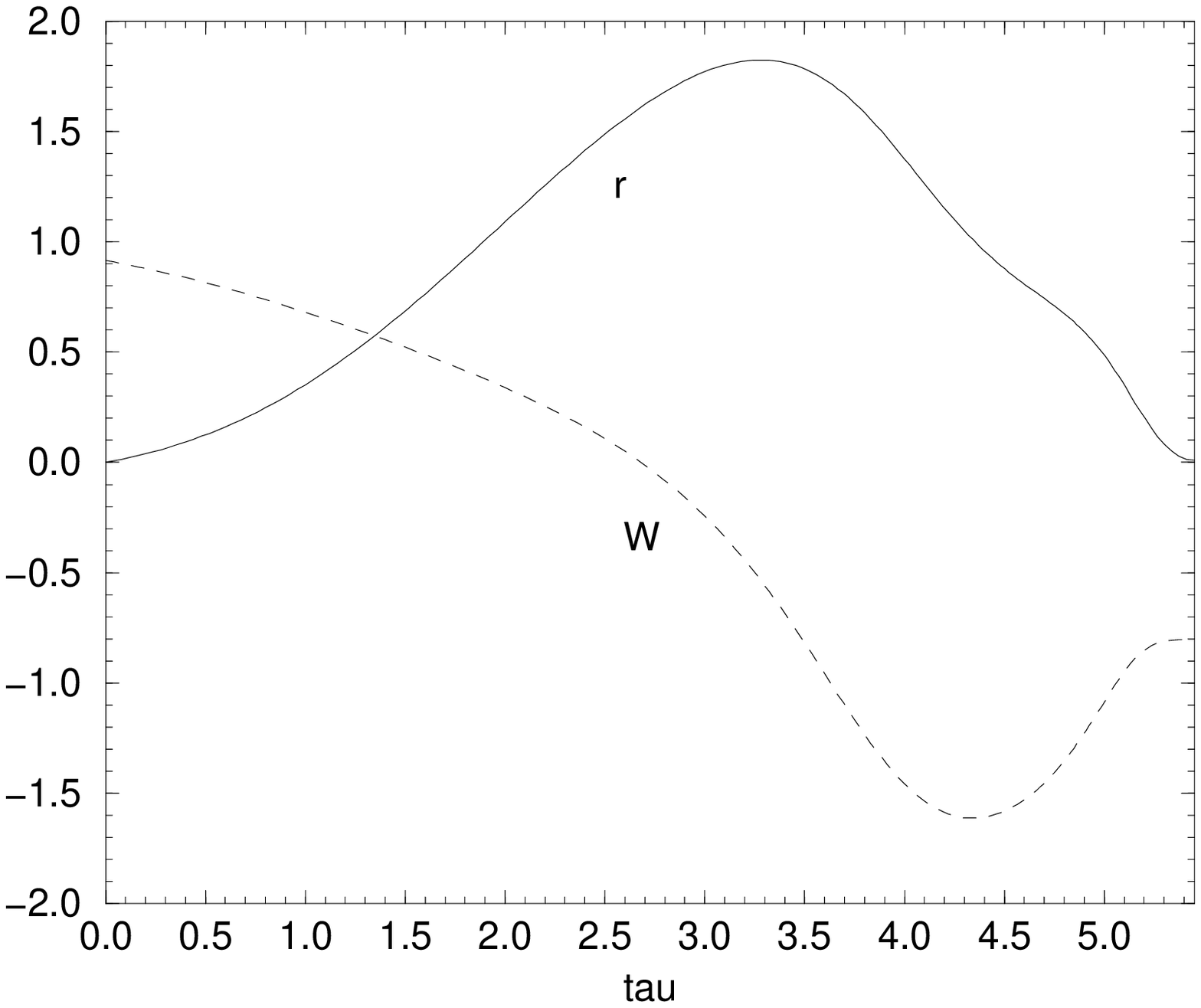,width=0.48\hsize,%
      bbllx=1.8cm,bblly=1.5cm,bburx=20.0cm,bbury=20.0cm}
  }
\labelcaption{naprn34}{
Two different NAPRN solutions with one zero of $W$ between $r=0$ and $r=r_h$.}
\end{figure}

\newpage

\begin{figure}
\hbox to\hsize{\hss
   \epsfig{file=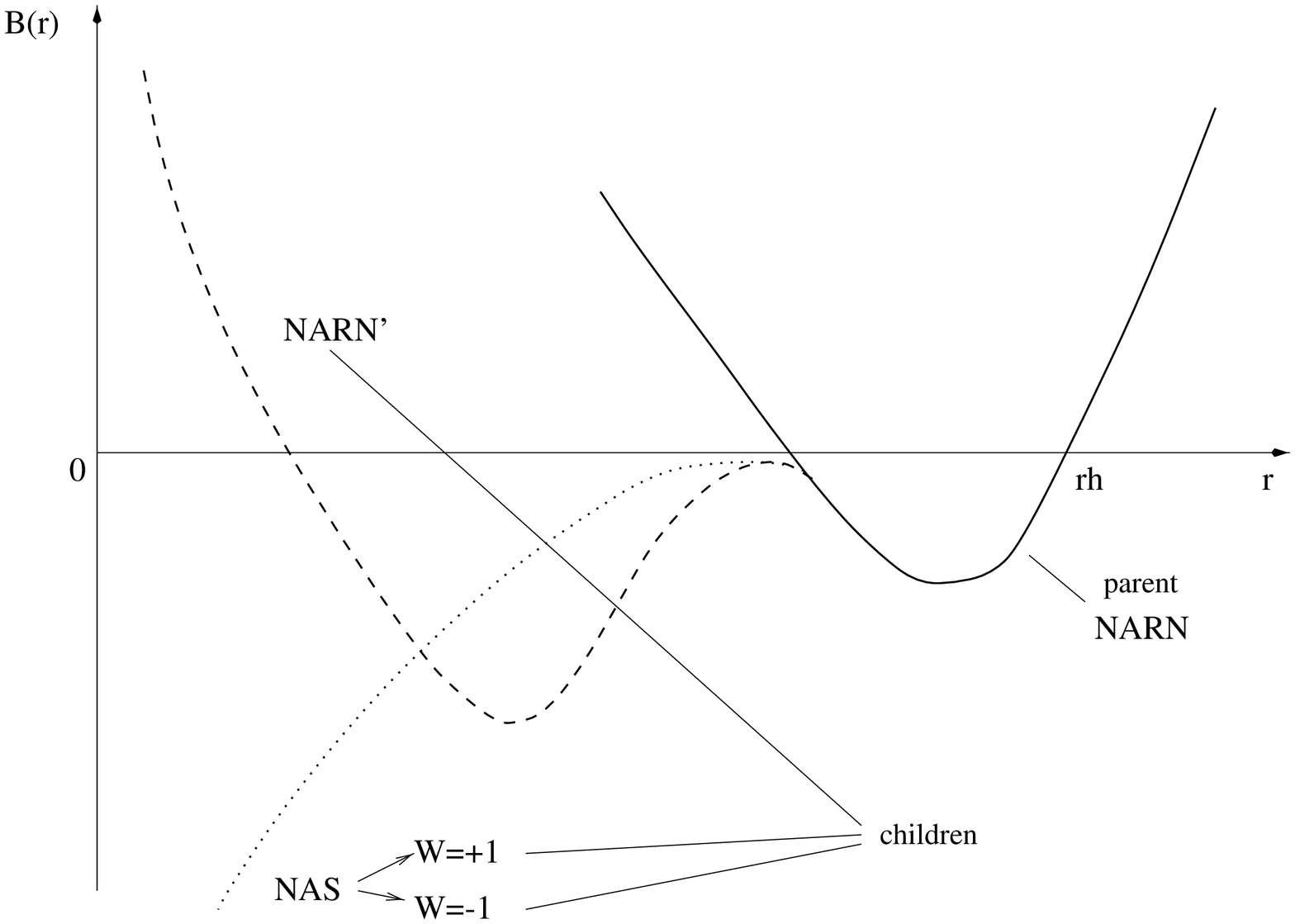,width=\hsize,%
   bbllx=1.3cm,bblly=0.0cm,bburx=20.0cm,bbury=19.8cm}%
  }
\labelcaption{children}{
Schematic view of the NARN solutions and their children.}
\end{figure}
\end{document}